# Nonequilibrium mesoscopic conductance fluctuations as the origin of 1/*f* noise in epitaxial graphene

C.-C. Kalmbach[1*], F. J. Ahlers[1†], J. Schurr[1], A. Müller[1], J. Feilhauer[1,3], M. Kruskopf[1], K. Pierz[1], F. Hohls[1], and R. J. Haug[2]

[1]*Physikalisch-Technische Bundesanstalt, Bundesallee 100, 38116 Braunschweig, Germany*
[2]*Institut für Festkörperphysik, Leibniz Universität Hannover, 30167 Hannover, Germany*
[3]*Institute of Electrical Engineering, Slovak Academy of Sciences, 84104 Bratislava, Slovakia*

We investigate the 1/*f* noise properties of epitaxial graphene devices at low temperatures as a function of temperature, current and magnetic flux density. At low currents, an exponential decay of the 1/*f* noise power spectral density with increasing temperature is observed that indicates mesoscopic conductance fluctuations as the origin of 1/*f* noise at temperatures below 50 K. At higher currents, deviations from the typical quadratic current dependence and the exponential temperature dependence occur as a result of nonequilibrium conditions due to current heating. By applying the theory of Kubakaddi [S. S. Kubakaddi, Phys. Rev. B 79, 075417 (2009)], a model describing the 1/*f* noise power spectral density of nonequilibrium mesoscopic conductance fluctuations in epitaxial graphene is developed and used to determine the energy loss rate per carrier. In the regime of Shubnikov-de Haas oscillations a strong increase of 1/*f* noise is observed, which we attribute to an additional conductance fluctuation mechanism due to localized states in quantizing magnetic fields. When the device enters the regime of quantized Hall resistance, the 1/*f* noise vanishes. It reappears if the current is increased and the quantum Hall breakdown sets in.

## 1. INTRODUCTION

Graphene, a monolayer of carbon atoms in a hexagonal lattice, is a promising material for a variety of electronic applications[1,2,3]. Low-frequency noise, also referred to as flicker noise or 1/*f* noise, is a common phenomenon caused by various physical mechanisms and found in numerous systems[4,5], including electronic transport in graphene devices[6-9]. At low temperature the noise properties of graphene devices are of particular interest for its application in metrology as a quantum Hall resistance[10-13] and impedance[14] standard. In low-temperature diffusive transport in a disordered conductor like graphene, quantum interference effects arise due to phase-coherent transport of electrons. When the phase-coherence length $L_\Phi$ of the electronic wave function is much longer than the elastic mean free path *l*, the effect of weak localization[15-18] occurs. It results from the interference of charge carriers backscattered from impurities along clockwise and counterclockwise paths and appears as a reduction of the average conductance at zero magnetic flux density.

Another consequence of quantum coherence in diffusive conductors is the phenomenon of universal conductance fluctuations (UCF)[19,20], which arise from interference of phase-coherent charge carriers between all possible paths through the device. UCF are sample-specific and occur as a function of magnetic flux density, chemical potential, or impurity configuration because these parameters change the paths of charge carriers through the device. When the size of the device *L* is smaller than the phase-coherence length ($L < L_\Phi$), then the amplitude of the conductance fluctuations is of the universal magnitude $e^2/h$, independent of the device size and the degree of disorder. For large devices ($L \gg L_\Phi$), the amplitude of the conductance fluctuations decays with increasing device size or decreasing phase-coherence length and is no longer of the order of $e^2/h$, but smaller due to ensemble averaging. In this case the conductance fluctuations are often referred to as mesoscopic conductance fluctuations (MCF).

In metals[19,20] and conventional semiconductors[21,22,23], an increase in temperature causes a power-law decay in amplitude of the mesoscopic conductance fluctuations, resulting from a temperature-induced dephasing due to the thermal energy $k_B T$. In graphene, the temperature dependence of MCF remains subject to controversy as Skákalová et al.[24] and Rahman et al.[9] observe different types of exponential decay and Bohra et al.[25] find a power-law decay.

However, these quantum conductance fluctuations (UCF and MCF) exhibit a strong sensitivity to impurity motion.[19] Consequently, a temporal fluctuation of the impurity configuration results in time-dependent quantum conductance fluctuations, which give rise to noise with a characteristic 1/*f*-type power spectral density (PSD).[19,9,26,27] Therefore, in large devices, as used in metrological applications of graphene, the existence of MCF can be probed through the investigation of 1/*f* noise.

For the usually applied very low current densities in noise measurements, a quadratic current dependence of the power spectral density $S_I$ indicates conductance fluctuations in equilibrium conditions as the origin of $1/f$ noise. On the other hand, in nonequilibrium conditions with electron-electron interaction taken into account, Ludwig et al.[28] predict the variance of mesoscopic conductance fluctuations to be inversely proportional to the applied voltage $V$: $\langle \delta G^2 \rangle \sim 1/V$. This would result in a linear current dependence of $S_I$. Such behavior has been observed in manganite thin films[29] and high-$T_c$ superconducting cuprates[30]. In these publications, the linear current dependence is ascribed to the effect of weak localization.

Here, we investigate the low-temperature $1/f$ noise properties of large-area epitaxial graphene devices in the absence of a magnetic field, in the regime of Shubnikov-de Haas oscillations, and in the regime of quantized Hall resistance. An exponential decay of the $1/f$ noise power spectral density with increasing temperature is observed at low currents, indicating mesoscopic conductance fluctuations as the origin of $1/f$ noise. This supports the observation by Skákalová et al.[24] and contributes to the controversial discussion about the temperature dependence of mesoscopic conductance fluctuations in graphene[9,25]. At low temperature and relatively high current (as required for metrological precision measurements of the quantized Hall resistance), we find a nonquadratic current dependence of the noise power spectral density $S_I$ and a nonexponential temperature dependence of $S_I$. We interpret the observed phenomena in terms of mesoscopic conductance fluctuations, which exhibit nonequilibrium conditions due to current heating effects at high currents. This allows us to determine the energy loss rate per carrier from the temperature and current dependence of the $1/f$ noise power spectral density. Furthermore, $S_I$ strongly increases when a magnetic field is applied and the device enters the regime of Shubnikov-de Haas oscillations. At the peak of the longitudinal resistance (corresponding to half-filling of the Landau level $n = 1$ in our device), the PSD reaches a maximum and the current dependence becomes linear. We find that in this regime an additional conductance fluctuation mechanism dominates, which can be related to localized states in the quantum Hall regime. Finally, when the device enters the quantized Hall regime, which is signified by vanishing longitudinal resistance, also the $1/f$ noise vanishes, since conductance quantization and noiseless current are inseparable.

## 2. EXPERIMENTAL METHODS

The measurements were carried out on large-area Hall bar devices (400 μm x 100 μm) patterned by electron beam lithography on epitaxial graphene grown on the silicon-terminated face of a 6H silicon carbide substrate. An optical microscope image of such a device is shown in FIG. 1(a). The graphene film had been grown in argon at atmospheric pressure within 5 minutes at a temperature of 1750 °C[31]. Raman spectroscopy was used, on samples with similar growth conditions as the one presented here, to confirm the presence of monolayer graphene, with only very few, electrically separated bilayer patches (FIG. 1(b)). Contrast-enhanced optical microscopy[32] and scanning-electron microscopy have confirmed similar film properties on the samples used here. Stable and low-resistance contacts ($< 10\ \Omega$ in the quantum Hall regime) were fabricated by an optimized two-step metallization process, Ti/Au (10 nm/30 nm) followed by Au (50 nm), which enables direct contact between the second gold layer and the graphene edge[33]. Photochemical gating[34] was applied to reduce the charge-carrier concentration by covering the sample with two polymers (70 nm of PMMA resist followed by 300 nm of ZEP520A resist) and subsequent UV radiation at room temperature. The main results were confirmed on three different devices. In this paper we present data from just one particular device, except for the data in FIG. 9.

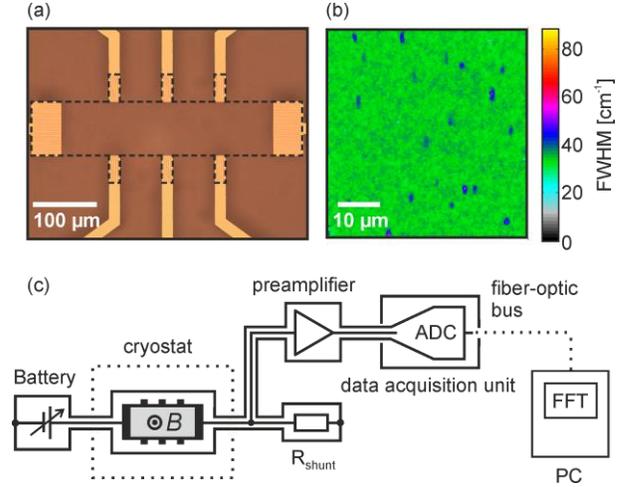

FIG. 1. (Color online) (a) Optical micrograph of the large-area epitaxial graphene Hall bar device used in the measurements presented in this paper. (b) Map of the Raman 2D-peak width of a sample with similar growth conditions shows monolayer graphene (green) with only very few, electrically separated bilayer patches (blue). (c) Low-frequency noise measurement setup.

All measurements were performed with the device in a cryomagnetic $^3$He system with a superconducting solenoid and coaxial measuring leads. Four-terminal DC measurements of the Hall resistance $R_{xy}$ and the longitudinal resistance $R_{xx}$, as well as two-terminal DC measurements of the source-drain resistance $R_{sd}$, were carried out by a scanning voltmeter while the source-drain current $I$ was provided by a battery-operated current source. Precision DC measurements of the longitudinal resistance and the quantized Hall resistance at the $v = 2$ plateau were performed in four-terminal configuration using a cryogenic current comparator bridge with the graphene device measured against a well-known GaAs quantum Hall device at a DC source-drain current of $I = 30$ µA.

FIG. 1(c) shows a schematic drawing of the noise measurement setup. Low-frequency noise measurements were carried out in a two-terminal-pair configuration by applying a constant voltage, provided by a low-noise battery, between the source contact and the outer conductor. The current fluctuations caused by the device were obtained by measuring the noise of the voltage drop across a 12.9 kΩ shunt resistor, which was connected to the drain contact of the device and the outer conductor and was cooled to 4 K in a liquid helium storage dewar. The measured voltage fluctuations were amplified by a low-noise preamplifier[35] (equivalent rms input noise voltage: 0.5 nV/√Hz) and recorded by an analog-to-digital converter model PXI-4461 from National Instruments[36] at a scan rate of 20000 Hz. Each trace of voltage fluctuations was recorded for 2 seconds and subsequently Fourier-transformed to obtain a spectrum in the frequency range of 20 Hz to 5000 Hz. 50-100 single spectra were averaged to reduce statistical errors. For all noise measurements, the intrinsic noise of the measurement setup and the Johnson-Nyquist noise of the device at each temperature, obtained at zero battery voltage and averaged over 200 spectra, was subtracted (see supplemental material). The resulting spectra of the frequency-dependent $1/f$ noise power spectral density are fitted by $S_I(f) = S_{I0} \cdot (f_0/f)^\alpha$ with the fitted noise power spectral density $S_{I0}$ at the reference frequency $f_0$ and the frequency exponent α. To characterize a whole spectrum by just one number, we quote in the following the fitted noise power spectral density $S_{I0}$ at $f_0 = 80$ Hz.

## 3. RESULTS AND DISCUSSION

In this section the experimental results are presented and discussed in three subsections. First, magnetotransport measurements characterizing the electrical transport properties of the device are presented. Next, the temperature and current dependences of the $1/f$ noise power spectral density are presented and interpreted in terms of nonequilibrium mesoscopic conductance fluctuations. In the last subsection, the $1/f$ noise properties in quantum Hall plateau transitions are studied and additional conductance fluctuation mechanisms in high magnetic fields are discussed.

### A. Electrical DC magnetotransport measurements

Electrical characterization by magnetotransport measurements is performed at a bath temperature of $T = 0.4$ K and a DC source-drain-current of $I = 10$ µA. An electron concentration of $n = 4.08 \cdot 10^{11}$ cm$^{-2}$ is derived from the slope of the Hall resistance around zero magnetic flux density (dashed line in FIG. 2). An electron mobility of $\mu = 7426$ cm$^2$/Vs is determined from $n$ and $R_{xx}$ at $B = 0$. The electron mean free path $l$ is calculated from the mobility to be about $l ≈ 55$ nm. Therefore, the electrical transport is in the diffusive regime ($L, L_\Phi \gg l$), regarding the large device size and typical coherence lengths of several hundred nanometers in epitaxial graphene on SiC with similar mobility and carrier concentration in this temperature range.[18] DC precision measurements of the quantized Hall resistance and the longitudinal resistance revealed an accurate quantization of the quantum Hall plateau at $\nu = 2$ with metrological precision of several parts in $10^9$ as well as a vanishing longitudinal resistance at magnetic flux densities $B > 7.5$ T for $I = 30$ µA. Note that the well quantized Hall resistance at $\nu = 2$ is a further evidence for the presence of largely single-crystalline monolayer graphene in our devices. Both the longitudinal resistance $R_{xx}$ and the Hall resistance $R_{xy}$ contribute to the two-terminal source-drain resistance $R_{sd}$, which is the resistance relevant to the two-terminal-pair noise measurements described below.

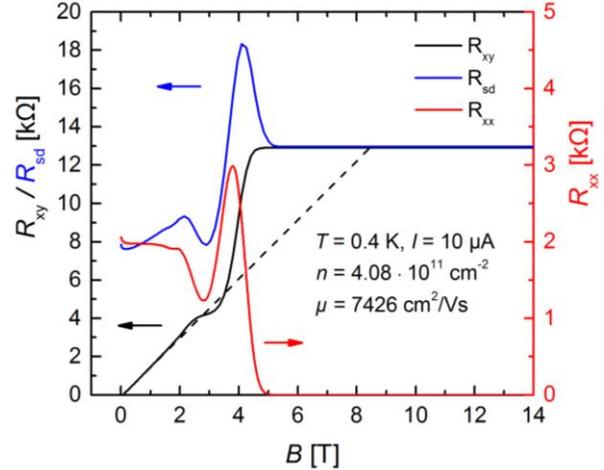

FIG. 2. (Color online) Hall resistance (black), longitudinal resistance (red) and two-terminal source-drain resistance (blue) as a function of magnetic flux density. The device had been tuned to the carrier concentration used in all noise measurements presented below.

### B. Temperature dependence of $1/f$ noise

Exemplary noise spectra measured for $I = 10$ µA at different temperatures and zero magnetic flux density are shown in FIG. 3 on a double-logarithmic scale. For all temperatures we find a typical power-law dependence of $1/f^\alpha$ with exponents $0.9 < \alpha < 1$ for the noise power spectral density $S_I$, in agreement with $1/f$-type noise. This is also the case at all measured magnetic flux densities discussed in section C.

FIG. 4 shows the temperature dependence of the fitted noise power spectral density $S_{I0}$ at a reference frequency of $f_0 = 80$ Hz for temperatures between $T = 0.4$ K and $T = 235$ K and a current of $I = 10$ µA.

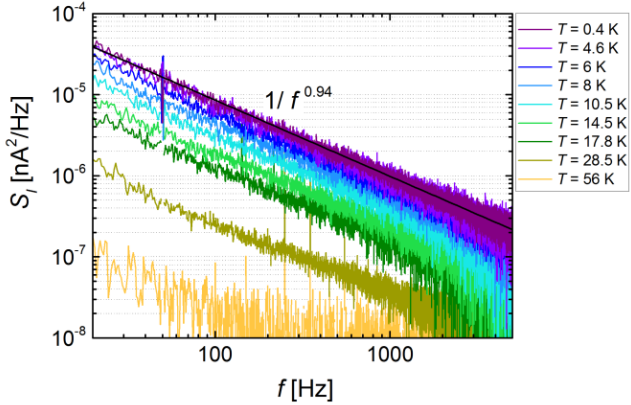

FIG. 3. (Color online) Frequency-dependent noise power spectral density $S_I$ as a function of frequency showing a $1/f$-type spectrum at all measured temperatures, zero magnetic flux density and a current of $I = 10$ µA. A power-law fit of $S_I$ at $T = 0.4$ K is shown by the black solid line.

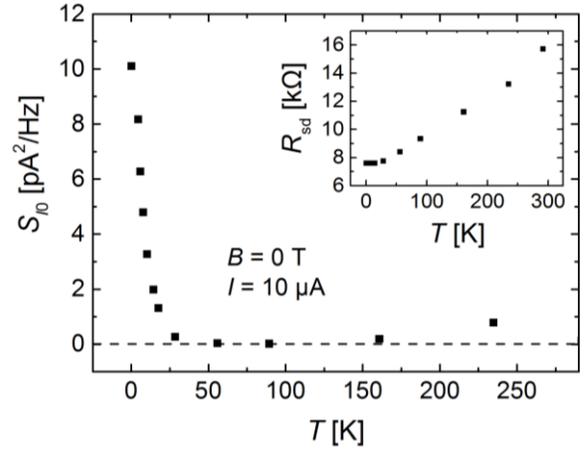

FIG. 4. Temperature dependence of the noise power spectral density $S_{I0}$ at 80 Hz for $I = 10$ µA at zero magnetic flux density. Below $T = 50$ K, $S_{I0}$ strongly increases with decreasing temperature, whereas the source-drain resistance of the device is approx. constant in this temperature range (inset).

At temperatures above $T = 56$ K, the noise power spectral density increases monotonically with temperature, predominantly due to thermal activation of mobile defects.[4,27,37] In this temperature range, the PSD exhibits a quadratic current dependence, as expected for conductance fluctuations, even at large currents. Also, the source-drain resistance $R_{sd}$ increases linearly with temperature, indicating phonon scattering as the origin of its temperature-dependence.[38]

Below $T \approx 50$ K, $S_{I0}$ strongly increases with decreasing temperature by more than two orders of magnitude. This increase is not related to the graphene source-drain resistance, which is approx. constant between $T = 0.4$ K and $T \approx 30$ K (see inset of FIG. 4).

The current dependence of $S_{I0}$ normalized to $S_{I0}(100$ µA) is shown in FIG. 5 in the temperature range below $T = 30$ K. At 28.5 K, the current dependence is still quadratic in the whole range of measured currents. A deviation from the quadratic current dependence occurs at very low temperatures. In the inset of FIG. 5 the exponent $b$ of the current dependence, obtained from a power-law fit $S_{I0} = a \cdot I^b$ in the current range 5-100 µA, is given as a function of temperature. The deviation from $b = 2$ is most distinct at the lowest temperature of $T = 0.4$ K and high currents. At currents below $I = 10$ µA, a deviation from a quadratic dependence is observed below $T \approx 6$ K, whereas at higher currents the onset temperature of this deviation increases. At 80-100 µA, $b$ deviates from 2 at temperatures below $T \approx 15$-20 K. At the lowest temperature of $T = 0.4$ K, a quadratic current dependence is not observed down to currents of $I = 0.25$ µA, but rather $b \approx 1.5$ is found (see supplemental material).

FIG. 6 shows the $1/f$ noise PSD data from FIG. 5 as a function of temperature (below 30 K) for various currents $I$. At low currents (e.g. $I = 5$ µA), the PSD decays exponentially with increasing temperature, and is best fitted by a $S_{I0} \sim \exp(-T/T_f)$ dependence as indicated by the red dot-and-dashed line. The exponential temperature dependence of the PSD specifically in this temperature range, together with a quadratic current dependence, strongly indicates mesoscopic conductance fluctuations as the origin of the $1/f$ noise. An exponential temperature decay was also observed for universal/mesoscopic conductance fluctuations in graphene in other studies[24,9] and differs from the power-law dependence usually found in normal metals in the phase-coherent regime[39]. However, the exact exponential behavior of the temperature dependence of mesoscopic conductance fluctuations is still controversial. Skákalová et al. find an $\exp(-T/T_f)$-dependence, which is in agreement with our measurements, whereas Rahman et al. observe an $\exp(1/T)$-dependence.

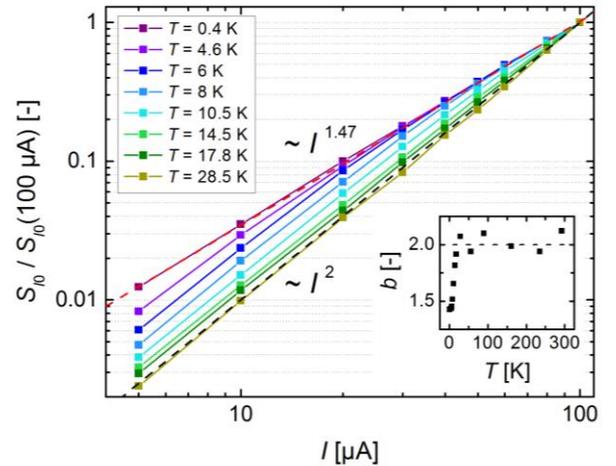

FIG. 5. (Color online) Current dependence of $S_{I0}$ normalized to $S_{I0}(100$ µA) at various temperatures and $B = 0$. The black dashed line indicates a quadratic current dependence. At very low temperatures and high currents, a deviation from the quadratic current dependence is observed, as illustrated by the red dashed line with $b = 1.47$ for $T = 0.4$ K. Solid lines connecting the points are a guide to the eye. The inset shows the exponent of the current dependence $b$ as a function of temperature.

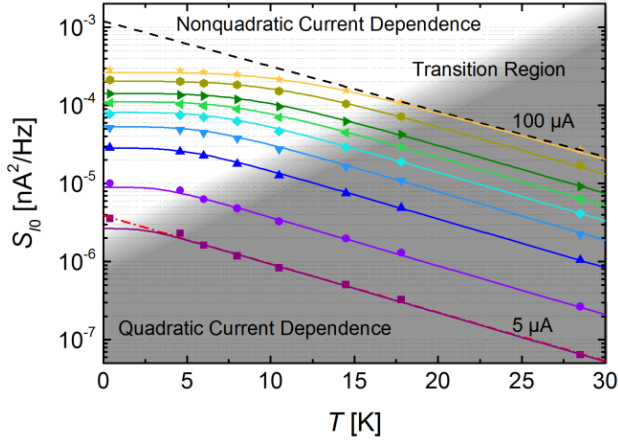

FIG. 6. (Color online) Temperature dependence of the noise power spectral density $S_{I0}$ at 80 Hz for $B = 0$, temperatures below $T = 30$ K and currents of 5 µA, 10 µA, 20 µA, 30 µA, 40 µA, 50 µA, 60 µA, 80 µA, and 100 µA. Symbols correspond to the measured data. The red dot-and-dashed line as well as the black dashed line indicate an exponential temperature dependence of $S_{I0} \sim \exp(-T/T_f)$. The solid lines are simultaneously fitted to all shown data points according to the theory of current heating by Kubakaddi[41], combined with the empirically found $\exp(-T/T_f)$-dependence. The grey background marks a region with quadratic current dependence of $S_{I0}$, whereas a nonquadratic current dependence is indicated by a white background.

At high currents and below approx. 6-15 K, the PSD deviates from the exponential temperature dependence (which is indicated by the black dashed line for $I = 100$ µA) and saturates at the lowest temperatures. This nonexponential decay with temperature is related to an exponent of the current dependence smaller than 2, as described before, with values down to $b = 1.42$ at $T = 0.4$ K and $I = 80$-$100$ µA. Points marked with grey background in FIG. 6 are consistent with a quadratic current dependence, whereas a white background corresponds to the regime of nonquadratic current dependence.

The following interpretation of our noise measurements in the framework of mesoscopic conductance fluctuations is further corroborated by the observation of residual conductance fluctuations in the DC measurements of $R_{xx}$ at low temperature and intermediate magnetic flux density (see supplemental material).

We explain the observed features of the low-temperature noise power spectral density $S_I$ in epitaxial graphene as follows.

Quantum interference effects arise at low temperatures due to an increased phase coherence length $L_\Phi$. As the carrier concentration of our graphene is not perfectly homogeneous and is affected by charged impurities, surface contaminations and substrate/buffer layer inhomogeneity, different current paths around such imperfections can interfere with each other as a result of the phase coherence. This appears as mesoscopic conductance fluctuations when the interference pattern changes as a function of, e.g., magnetic flux density or chemical potential. In large area devices, as used for metrological applications, these fluctuations are barely visible in the longitudinal resistance as MCF are averaged with larger device size.

Nevertheless, due to the strong sensitivity to impurity motion, the MCF manifest themselves in the 1/f noise of the device as a result of the temporal fluctuation of the impurity configuration. In disordered metals, MCF theory explains an enhancement of 1/f noise at low temperature as a result of an increased sensitivity of the conductance to impurity motion.[27,39] As our measurements show, this is also the case for epitaxial graphene and gives rise to the strong increase of 1/f noise at temperatures below $T \approx 50$ K (FIG. 4).

In the low current limit where current heating is negligible, the MCF-induced noise power spectral density $S_I$ is proportional to the square of the current and decays exponentially with increasing temperature[24], expressed mathematically by

$$S_I \sim I^2 e^{-\frac{T_e}{T_f}} \quad (1)$$

where $T_e$ is the electron temperature and $T_f$ is the exponential decay parameter. In this model, the 1/f noise decays with increasing temperature because the thermal energy begins to destroy the phase coherence, which is essential to quantum interference.

Currents of up to several tens of microamperes are required for metrological precision measurements of the quantized Hall resistance. At these high current levels, current heating must be considered since the electron temperature $T_e$ may be elevated from the lattice temperature $T$.[40] Using the theory of current heating in graphene by Kubakaddi[41], the effective minimum electron temperature $T_e$ can be derived from the energy loss rate per carrier $F(T)$, which is given by Eqs. (16, 20) in [41]. From those equations one calculates $T_e$ as

$$T_e = \sqrt[4]{\frac{I^2 R_{xx}}{\alpha n A} + T^4} \quad (2)$$

where $A$ is the area of the graphene sheet and $\alpha$ is the coefficient of the $T^4$-dependent energy loss rate per carrier.

Combining Eq. (1) and Eq. (2) leads to a formula which describes the measured noise power spectral density as a function of the current $I$, taking into account the effect of current heating and thermal destruction of phase coherence:

$$S(I,T) = aI^2 e^{-\sqrt[4]{\left(\frac{I}{I_0}\right)^2 + \left(\frac{T}{T_f}\right)^4}} \quad (3)$$

with $a$ parameterizing the amplitude of the noise power and

$$\frac{1}{I_0^2} = \frac{1}{T_f^4} \frac{R_{xx}}{\alpha n A} \quad (4)$$

being introduced for convenience.

A least squares fit of Eq. (3) simultaneously to all experimental data in FIG. 6 is plotted as solid lines and yields the following values: $a = (1.6 \pm 0.03) \cdot 10^{-7}$ Hz$^{-1}$, $I_0 = (31.2 \pm 1.2)$ µA, $T_f = (6.95 \pm 0.06)$ K. The good quality of the fit supports the validity of our approach. At high currents and low temperatures, the model explains the observed nonexponential temperature dependence and the nonquadratic current dependence of $S_I$ as a result of the elevated electron temperature. This strongly indicates that hot-electron effects lead to nonequilibrium conditions of the mesoscopic conductance fluctuations which govern the strength of the 1/$f$ noise at high current densities and low temperatures. Note that the current dependence observed at the lowest temperatures is just apparently nonquadratic as a result of the nonequilibrium conditions causing an increase in electron temperature. Another way of saying the same is that, according to Baker et al.[18], $L_\Phi$ saturates at low temperature as a consequence of the finite electron temperature due to current heating. Consequently, $S_I$ saturates due to a saturation of $L_\Phi$ at low temperature.

Baker et al.[18] as well as Lara-Avila et al.[42] report about hot-electron effects in measurements of weak localization and of the phase-coherence length $L_\Phi$ in large-area epitaxial graphene down to very low current densities ($j \approx 1.5 \cdot 10^{-6}$ A/m) at low bath temperatures ($T = 10$ mK). In agreement with their findings, an effective minimum electron temperature of $T_e \approx 0.62$ K is calculated from Eq. (2) for our minimum current density of $j = 0.0025$ A/m at $T = 0.4$ K, already indicating nonequilibrium conditions in the electron system of our device. Therefore, the exponent of the current dependence is significantly smaller than 2 under these conditions (see supporting material). However, Rahman et al.[9] report about a quadratic current dependence of $S_I$ in their experiments at a much higher current density of $j \approx 0.14$ A/m and temperatures down to $T = 0.25$ K, ruling out current heating in their small exfoliated graphene devices. As mentioned above, Ludwig et al.[28] predict an effect of electron-electron interaction on the current dependence of mesoscopic conductance fluctuations via nonequilibrium conditions. Since small device size sets a low-energy cutoff to electron-electron interaction[43], the quadratic current dependence observed by Rahman et al.[9] at $j \approx 0.14$ A/m could be a result of their small device size.

However, in our large-area devices electron-electron interaction is present at zero and intermediate magnetic flux densities (see supporting material) and might be related to the heating of the electron system in our case.

Note that, since the current heating depends on the energy loss rate per carrier, this quantity can directly be extracted from our noise data. Usually, the energy loss rate per carrier is obtained from the current and temperature dependence of the Shubnikov-de Haas oscillation amplitude, or from an analysis of the temperature dependence of the weak localization peak around zero magnetic flux density. Determining it from the current and temperature dependence of the 1/$f$ noise power spectral density is therefore an alternative method for measuring the energy loss rate in graphene Hall bar devices. Based on the current heating model by Kubakaddi, an energy loss coefficient of $\alpha = (21.0 \pm 0.9) \cdot 10^{-18}$ WK$^{-4}$ is obtained from Eq. (4) by using the fit values for $I_0$ and $T_f$ as well as the area $A$, carrier concentration $n$ and longitudinal resistance $R_{xx}$ of our Hall bar device. This value of $\alpha$ is of the same order of magnitude as the theoretical prediction from Kubakaddi[41] for a device of similar carrier concentration and a deformation potential of $D = 19$ eV and the value of $\alpha$ extracted from weak localization data by Baker et al.[40]. Therefore, our results are another confirmation of the large energy loss rate per carrier in graphene by an independent measurement method. Since the larger energy loss rate per carrier is one of the factors for the enhanced breakdown currents observed in graphene-based quantum resistance standards, when compared to GaAs[40], this issue deserves further attention.

It should further be noted that from the energy loss rate coefficient α experimentally determined here as well as in [40], a value for the deformation potential $D$ of graphene can be extracted within the framework of the Kubakaddi theory. Calculating $D$ from our data yields a value of 25.8 eV. This is just at the upper end of the range of 10..30 eV quoted in literature[44,45]. All these values are, however, significantly higher than the value of (~4 eV) predicted from first-principles calculations[46]. While our result is yet another confirmation of the statement made elsewhere[47] that the experimentally determined values are systematically higher, we can give no explanation for this discrepancy.

### C. Magnetic field dependence of 1/$f$ noise

The magnetic field dependence of the noise power spectral density $S_{I0}$ at the reference frequency of $f_0 = 80$ Hz, $T = 0.4$ K and $I = 10$ µA is displayed in FIG. 7. The magnetic field dependence shows that the noise is qualitatively correlated with the longitudinal resistance $R_{xx}$. This can be seen by the peak of $S_{I0}$ around zero magnetic flux density, which originates from the weak localization peak in the longitudinal resistance[9], by the maximum of $S_{I0}$ around $B \approx 4$-5 T (corresponding to the Shubnikov-de Haas peak of the first Landau level), and by the vanishing of the noise PSD in the quantum Hall plateau region.

First, we investigate the peak of the $1/f$ noise PSD resulting from weak localization around zero magnetic flux density. A reduction of $S_{I0}$ by a factor of 1.55 is observed between $B = 0$ T and $B = 0.2$, in agreement with the findings of Rahman et al.[9] obtained at $T = 0.4$ K. This is another confirmation of our assumption that quantum interference noise due to mesoscopic conductance fluctuations is the source of the observed noise phenomena around $B = 0$ T. For very low temperatures $T \to 0$, UCF theory predicts a reduction of the $1/f$ noise PSD with increasing magnetic field by a factor of 2 (for low magnetic field < 0.5 T) in conventional systems with Landau-level quantization, because the cooperon contribution is gradually reduced while the diffusion contribution stays constant[48,49]. In contrast to this behavior in metals and conventional semiconductors, Rahman et al.[9] report a factor-of-4 reduction for graphene at a temperature of 250 mK and ascribe it to symmetry breaking between valley degrees of freedom. This reduction factor is temperature dependent and decreases to 1.65 at 0.4 K, comparable to the factor of 1.55 observed here.

We now turn to the increase of the $1/f$ noise with increasing magnetic field, showing a maximum at $B = 4.4$ T. The large increase of the noise by two orders of magnitude, particularly in the transition between two quantum Hall plateaus, cannot be explained by an increase in resistivity. In fact, the model described above in Eq. (3) would result in a decrease of the $1/f$ noise PSD as the resistivity increases, due to a stronger effect of current heating and the concomitant exponential suppression of noise with increasing electron temperature. Moreover, since the cyclotron radius $r_c = (\hbar/eB)\sqrt{\pi n}$ becomes smaller than the elastic mean free path of $l \approx 55$ nm above 1.4 T, the conductivity fluctuations are no longer well described by the conventional theory of universal conductance fluctuations.[50,51]

Therefore, we suggest that in the quantum Hall transition regime a second conductance fluctuation mechanism becomes dominating, which is described by Machida et al.[52] as the result of a network of compressible and incompressible subregions[53]. They found that in quantum Hall transitions the phase coherence is not significant in determining the fluctuation pattern even though the conductor is in a coherent regime. Moreover, in the presence of carrier concentration inhomogeneities across the device, the energy $E_N$ of the $N$th Landau level fluctuates in space around the Fermi energy $E_F$ (FIG. 8). Consequently, if the average filling factor of Landau levels $v$ takes a noninteger value, the two-dimensional electron system splits into nonconducting incompressible subregions with local filling factors of $v = N$ and $v = N - 1$ (and $v = N+1$, if the potential inhomogeneity is larger than the energy splitting between the Landau levels, as it has been reported for graphene devices[54]). These subregions are separated by a percolating conductive network of compressible stripes with $E_N = E_F$, whose topology significantly affects the conductance of the device.[52] Hence, irregular time-dependent fluctuations of the impurity configuration will modify the local carrier concentration as well as the topology of the network of compressible stripes and consequently result in strong $1/f$ noise of the conductance in the corresponding magnetic field regions.

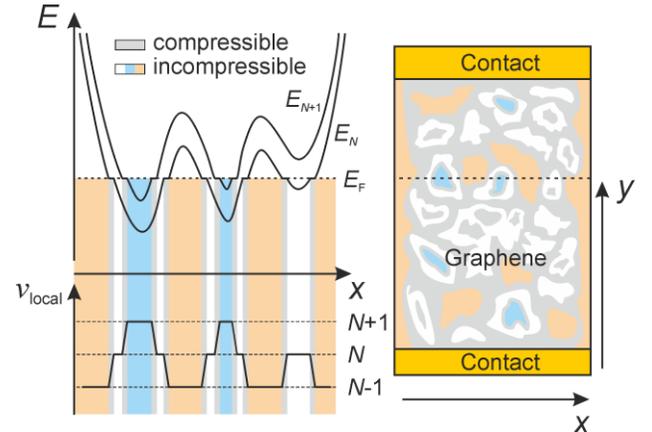

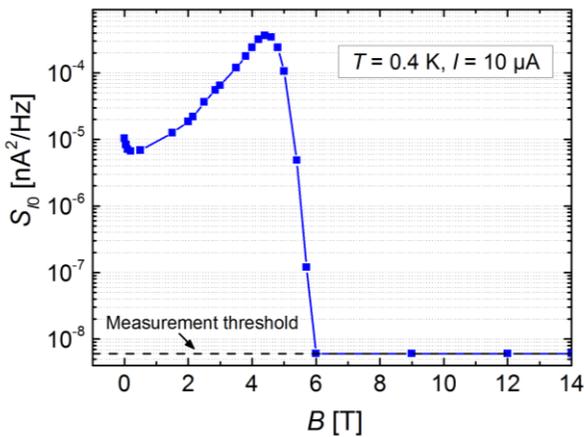

FIG. 7. (Color online) Noise power spectral density $S_{I0}$ at 80 Hz as a function of magnetic flux density at $T = 0.4$ K and $I = 10$ µA. $S_{I0}$ strongly increases in the regime of Shubnikov-de Haas oscillations. When the device enters the quantized Hall state, the noise decreases below the measurement threshold of our setup.

FIG. 8. (Color online) Scheme of the potential landscape of a graphene quantum Hall device at the peak of a Shubnikov-de Haas oscillation (half filling of the $N$th Landau level). As a result of the potential inhomogeneity, the two-dimensional electron gas splits into incompressible subregions with different local filling factors (white, blue, sand-colored) as well as compressible subregions (grey). These compressible stripes form a percolating conductive network that significantly determines the conductivity of the device. Time-dependent fluctuations of the impurity configuration as well as charging events of compressible islands modify the conductivity of this network and result in $1/f$ noise of the conductivity.

This interpretation is supported by taking a closer look at the data in FIG. 2 and FIG. 7, which reveals that the maximum of the 1/$f$ noise PSD (at $B \approx 4.4$ T) occurs at larger magnetic flux density than the maximum of $R_{sd}$ (at $B \approx 4.1$ T) or $R_{xx}$ (at $B \approx 3.8$ T): On the low $B$ side of an $R_{xx}$-peak the potential landscape is formed by incompressible islands in the compressible Fermi sea[55], whereas on the high $B$ side of an $R_{xx}$-peak the landscape corresponds to compressible lakes on insulating, incompressible land. Here, on the high $B$ side, transport occurs by hopping or tunneling between the compressible lakes. This is more sensitive to fluctuations in the potential landscape, which explains why the noise PSD peak shifts to a higher magnetic field.

In a more detailed picture, confined compressible islands on a mesoscopic scale are localized states that effectively behave as quantum dots. Such quantum dot-like localized quantum Hall states have been associated with conductance fluctuations in the regime of quantum Hall transitions before[54,56,57]. They are the origin of line structures parallel to integer filling factors observed in the $n$-$B$-plane of (trans-)conductance[54,56,57] as well as of scanning single-electron-transistor measurements of GaAs/AlGaAs heterostructure[58] and graphene[59] QHE devices. The typical size of the compressible dots is found to be about 60-400 nm[54,58,59,60]. Fingerprints of Coulomb blockade physics in these states and the existence of a network of dots in graphene have been confirmed by Lee et al.[54], who observed the typical Coulomb diamonds near the $v = 0$ state. According to Lee et al.[54], there is a large difference $\Delta n \approx 3\text{-}4 \cdot 10^{11}$ cm$^{-2}$ between the extrema $n_{min}$ and $n_{max}$ of the spatial variation in carrier concentration in graphene, more than an order of magnitude larger compared to GaAs/AlGaAs heterostructures[58]. They investigated a small suspended graphene device, but for large epitaxial graphene devices this should be even more so. As a result, localized states with filling factors $v = N-1$ and $v = N+1$ are present when the Fermi energy corresponds to the $N$-th Landau level at the maximum of the longitudinal resistance in quantum Hall transitions. Charging of such quantum dot-like compressible islands alters the transport channels through the device. Therefore, in graphene, the dominant conductance fluctuation mechanism due to the landscape of compressible and incompressible subregions in the quantum Hall transition regime can result from the fluctuations in the network of compressible stripes, from charging of localized states, and transport across localized states in the bulk of the device.

As an additional mechanism, we suggest that the amplitude of conductance fluctuations generally increases with increasing magnetic flux density by considering the following argument. In quantum Hall transitions, scattering mainly occurs in the highest-energy occupied Landau level since transport through the device in the lower-energy fully occupied levels occurs in dissipationless edge channels. Thus, 1/$f$ noise due to conductance fluctuations in strong magnetic fields should originate from the highest-energy occupied Landau level. When the magnetic flux density is increased, that level is responsible for a larger fraction of the current transport since the number of occupied Landau levels decreases. Therefore, a larger fraction of the total current would be subject to conductance fluctuations which results in an increase of 1/$f$ noise power spectral density with increasing magnetic flux density in the regime of Shubnikov-de Haas oscillations. A similar argument was given by Main et al.[61] for resistance fluctuations in the quantum Hall regime due to resonant backscattering of electrons in narrow quantum wires. We believe that this reasoning is also applicable here.

In epitaxial graphene, surface-donor states in the underlying SiC substrate act as a charge reservoir in proximity to the two-dimensional electron gas[62,63]. In strong magnetic fields, the carrier concentration of epitaxial graphene varies with the magnetic flux density due to a charge transfer between the surface-donor states and the graphene. During our measurements of 1/$f$ noise at fixed magnetic flux density, the average carrier concentration remains constant. Nevertheless, the surface-donor states in the SiC substrate might act as scattering centers similar to charged impurities and surface contaminations on the top surface of the graphene layer. In this respect, temporal fluctuations of the surface-donor states would affect the conductivity by the same mechanisms as a temporal fluctuation of the impurity configuration does. Since it is not possible to distinguish between these scattering mechanisms on the basis of our data, comparison with corresponding 1/$f$ noise measurements in graphene made by exfoliation or chemical vapor deposition could clarify the role of the charge reservoirs in the underlying SiC substrate.

The temperature dependence of the noise power spectral density $S_{I0}$ at 80 Hz for various currents $I$ at the peak of the longitudinal resistance is shown in FIG. 9, similar to the data in FIG. 6 at zero magnetic flux density. As illustrated by the red dot-and-dashed line, the temperature dependence is exponential as well with the same decay parameter $T_f$ as at zero magnetic field (FIG. 6). This suggests that mesoscopic conductance fluctuations still play a role in the fluctuation mechanism in quantum Hall transitions[64]. On the other hand, the current and temperature dependences of the 1/$f$ noise power spectral density at the peak of the longitudinal resistance cannot be fitted satisfactorily according to Eq. (3) if a quadratic current dependence of the underlying fluctuation mechanism is assumed and $b = 2$ is fixed. Instead, taking $b$ as a free parameter results in a much better fit and yields values of $b \approx 1$ to $b \approx 1.2$. In addition, the data at the peak of the longitudinal resistance is not described by the fit as well as it is the case for $B = 0$ T, suggesting that Eq. (3) does not include all mechanisms that arise in strong magnetic flux densities. This provides further evidence for the presence of a conductance fluctuation mechanism due to a network of compressible and incompressible subregions in the regime of quantum Hall transitions.

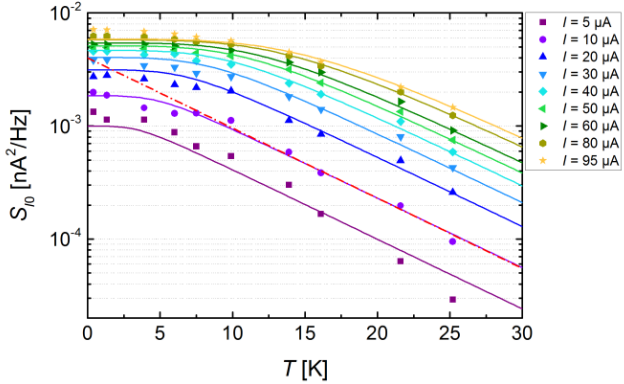

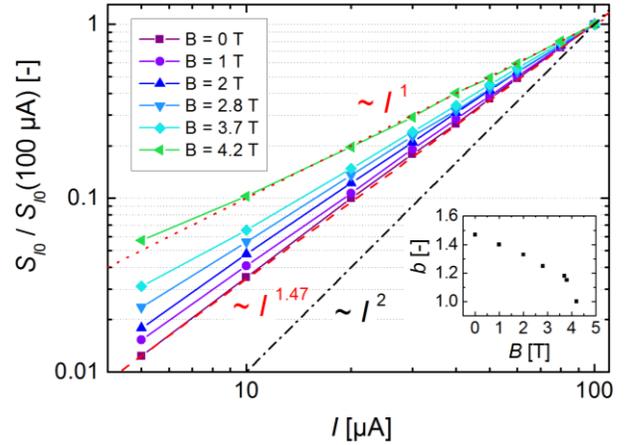

FIG. 9. (Color online) Temperature dependence of the noise power spectral density $S_{I0}$ at 80 Hz for various currents $I$, temperatures below $T = 30$ K and at the peak of the longitudinal resistance. Symbols correspond to the measured data. The red dot-and-dashed line indicates an exponential temperature dependence of $S_{I0} \sim \exp(-T/T_f)$, which also agrees well with the data at zero magnetic field. The decay parameter $T_f$ is the same for the red dot-and-dashed lines in FIG. 6 and in FIG. 9. The solid lines are a fit according to Eq. (3). Values of $b \approx 1$ to $b \approx 1.2$ are required to obtain a good fit, rather than a quadratic current dependence of the underlying fluctuation mechanism as it is the case for $B = 0$. Since this measurement is obtained from a similar device with slightly higher carrier concentration than all other measurements, the peak of the longitudinal resistance is at $B = 7.5$ T here.

The physical effects discussed above might also explain the magnetic field dependence of the $1/f$ noise PSD observed at temperatures of 80 K and 285 K by Rumyantsev *et al.*[65], since the quantum Hall effect in graphene persists up to room temperature in strong magnetic fields.

When the device enters the quantized Hall state, the noise decreases below the measurement threshold, as shown in FIG. 7 for $B \geq 6$ T. In the quantum Hall plateau region the noise vanishes due to a vanishing longitudinal resistance and a Hall resistance which is determined by fundamental constants to a very high precision, inhibiting fluctuations in the resistance of the graphene sheet. This demonstrates once again that the quantum Hall state is a macroscopic quantum state, which is robust, stable and does not allow for conductance fluctuations in the quantized Hall resistance. Both effects, the absence of current noise as well as the quantized Hall resistance are a manifestation of Fermi statistics and the Pauli principle. Therefore, conductance quantization and noiseless current are inseparable.[66] This requires the absence of backscattering, which is verified for our epitaxial graphene devices by the precision measurements of the quantized Hall resistance discussed before. This behavior was observed on GaAs/AlGaAs heterostructures quantum Hall devices[67,68,69] and is also verified for epitaxial graphene devices by our experiments. From the metrological point of view, this fact is an important prerequisite for low-noise precision measurements of the quantized Hall resistance. Furthermore, one could argue that the absence of $1/f$ noise to a certain level is another way of proving the resistance quantization in graphene to a certain, corresponding level: if the device still exhibits a finite longitudinal resistance then it will also show $1/f$ noise. This has been verified with a device which deviated from prefect quantization at $\nu = 2$ by only 2-3 parts in $10^9$, but still exhibited $1/f$ noise above our measurement threshold. Only the sensitivity of the noise measurement setup and the thermal noise of the quantized Hall resistance limit how accurate the resistance quantization can be verified by measuring the $1/f$ noise in graphene devices.

FIG. 10. (Color online) Current dependence of $S_{I0}$ normalized to $S_{I0}(100\ \mu A)$ for various magnetic flux densities at $T = 0.4$ K. The exponent $b$ of the current dependence decreases with increasing magnetic flux density (inset).

FIG. 10 shows the current dependence of $S_{I0}$ normalized to $S_{I0}(100\ \mu A)$ for various magnetic flux densities at the lowest measured temperature of $T = 0.4$ K. At zero magnetic flux density, the current dependence of $S_{I0}$ is nonquadratic due to current heating. With increasing magnetic flux density, the exponent $b$ of the current dependence decreases monotonously from $b = 1.47$ at $B = 0$ to $b = 1$ at $B = 4.2$ T. At the maximum of the source-drain resistance $R_{sd}$ ($B = 4.2$ T), the current dependence of the $1/f$ noise PSD becomes linear for this device, while other values $b < 1$ were observed for other devices. This behaviour might have two reasons. First, the nonquadratic current dependence could be attributed to current heating, as discussed at $B = 0$. Second, we discussed above that the dominating conductance fluctuation mechanism in strong magnetic fields might also exhibit an intrinsic current dependence smaller than $b = 2$.

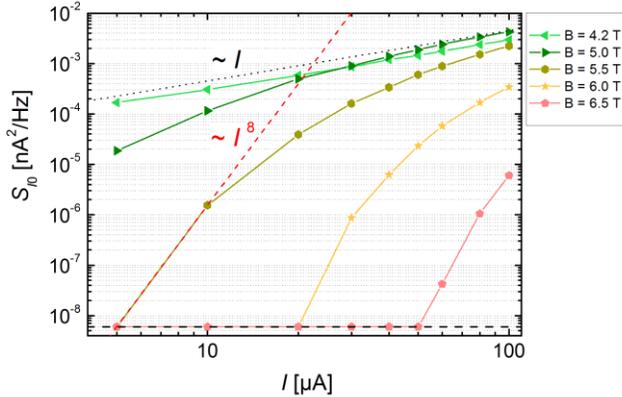

FIG. 11. (Color online) Current dependence of the 1/$f$ noise power spectral density $S_{I0}$ at 80 Hz in the QHE breakdown regime. At the onset of the breakdown, the exponent of the current dependence is larger than $b = 2$, as indicated by the red dashed line. At higher currents, the current dependence converges to $b < 2$. The black dashed line represents the measurement threshold.

Further increase of the magnetic flux density at low currents drives the device into the quantized Hall state (FIG. 11), accompanied by vanishing of the 1/$f$ noise. By increasing the current in the quantized Hall state, the device can be driven into the breakdown regime and the 1/$f$ noise sets in again. Here, the current dependence of the 1/$f$ noise power spectral density cannot be described by a single function $S_{I0} = a \cdot I^b$. In the breakdown regime, current heating is likely to occur in the device. According to our model described before, an elevated electron temperature should result in an exponent of the current dependence smaller than $b = 2$. Nevertheless, at the onset of the quantum Hall effect breakdown the exponent $b$ of the current dependence is much larger than $b = 2$, as indicated by the red dashed line in FIG. 11 that corresponds to $b = 8$. At higher currents, the current dependence seems to converge to some value $b < 2$. We interpret this behavior as follows. Two effects might contribute here to the strong current dependence of the 1/$f$ noise PSD. First, variable range hopping[70,71] needs to be considered at low temperatures and very low finite longitudinal resistivities in the quantum Hall regime. As a result of variable range hopping, the longitudinal resistivity is known to increase nonlinearly with temperature, which has been shown for GaAs[72] as well as for epitaxial graphene devices[73]. Furthermore, the electric field strength causes the same effect as an effective temperature in the variable range hopping mechanism[72,74]. Thus, the large increase of 1/$f$ noise PSD with increasing current could result from a strongly nonlinear current dependence of the longitudinal resistivity due to variable range hopping. Second, when the breakdown sets in, dissipation will arise first locally due to inhomogeneity of the carrier concentration and then spread over the entire area of the device with increasing source-drain current. Along with the dissipation, 1/$f$ noise will not extend over the whole device area at the first onset of the breakdown. Therefore, the 1/$f$ noise power spectral density shows a larger-than-quadratic increase at the onset of the breakdown, because with increasing current also the area affected by the breakdown will increase, and then converges asymptotically to $b < 2$ due to current heating.

## 4. CONCLUSIONS

In conclusion, we have studied the low-temperature 1/$f$ noise properties of epitaxial graphene devices as a function of temperature, current and magnetic flux density. We find an exponential decay of the noise power spectral density with increasing temperature, which indicates mesoscopic conductance fluctuations as the origin of 1/$f$ noise below 50 K. At high currents, the temperature dependence deviates from the exponential decay and the current dependence is nonquadratic, which both is a result of nonequilibrium conditions due to current heating. Using the theory of Kubakaddi, we develop a model for the 1/$f$ noise power spectral density of mesoscopic conductance fluctuations, which takes the effect of current heating into account. Based on our model, we calculate the energy loss rate per carrier for our device. This demonstrates a new method for determination of the energy loss rate per carrier in devices with 1/$f$ noise due to mesoscopic conductance fluctuations.

In the regime of quantum Hall transitions, the 1/$f$ noise power spectral density strongly increases. We suggest that a second conductance fluctuation mechanism, based on a network of compressible and incompressible subregions, dominates the 1/$f$ noise properties of epitaxial graphene under these conditions. In more detail, we attribute the conductance fluctuations in quantum Hall transitions to charging events of localized states in quantizing magnetic fields.

In the quantum Hall plateau region (at low currents), the 1/$f$ noise vanishes as a consequence of an accurately quantized Hall resistance, confirming that the absence of current noise and the quantized Hall resistance are inseparable. The 1/$f$ noise sets in again, when the current is increased and the quantum Hall effect breaks down. Therefore, we propose the measurement of 1/$f$ noise as an alternative way of proving the resistance quantization in graphene.

## ACKNOWLEDGEMENTS

This research has been performed within the EMRP project SIB51, GraphOhm. The EMRP is jointly funded by the EMRP participating countries within EURAMET and the European Union. We gratefully acknowledge the support by the Braunschweig International Graduate School of Metrology B-IGSM and NanoMet. The authors thank V. Bürkel for the technical support, as well as M. Götz and E. Pesel for DC precision measurements.

*cay-christian.kalmbach@ptb.de
†Franz-Josef.Ahlers@ptb.de

[1] K. S. Novoselov, A. K. Geim, S. V. Morozov, D. Jiang, Y. Zhang, S. V. Dubonos, I. V. Grigorieva, and A. A. Firsov, Science **306**, 666-669 (2004).
[2] K. S. Novoselov, A. K. Geim, S. V. Morozov, D. Jiang, M. I. Katsnelson, I. V. Grigorieva, S. V. Dubonos, and A. A. Firsov, Nature **438**, 197-200 (2005).
[3] K. S. Novoselov, V. I. Fal′ko, L. Colombo, P. R. Gellert, M. G. Schwab, and K. Kim, Nature **490**, 192-200 (2012).
[4] P. Dutta and P. M. Horn, Rev. Mod. Phys. **53**, 497 (1981).
[5] F. N. Hooge, T. G. M. Kleinpenning, and L. K. J. Vandamme, Reports on Progress in Physics **44**, 479-532 (1981).
[6] G. Liu, W. Stillman, S. Rumyantsev, Q. Shao, M. Shur, and A. A. Balandin, Applied Physics Letters **95**, 033103 (2009).
[7] A. N. Pal, S. Ghatak, V. Kochat, E. S. Sneha, A. Sampathkumar, S. Raghavan, and A. Ghosh, ACS Nano **5**, 2075-2081 (2011).
[8] A. A. Balandin, Nature Nanotechnology **8**, 549-555 (2013).
[9] A. Rahman, J. W. Guikema, and N. Marković, Phys. Rev. B **89**, 235407 (2014).
[10] A. Tzalenchuk, S. Lara-Avila, A. Kalaboukhov, S. Paolillo, M. Syväjärvi, R. Yakimova, O. Kazakova, T. J. B. M. Janssen, V. Fal'ko, and S. Kubatkin, Nature Nanotechnology **5**, 186-189 (2010).
[11] M. Woszczyna, M. Friedemann, M. Goetz, E. Pesel, K. Pierz, T. Weimann, and F. J. Ahlers, Applied Physics Letters **100**, 164106 (2012).
[12] A. Satrapinski, S. Novikov, and N. Lebedeva, Applied Physics Letters **103**, 173509 (2013).
[13] R. Ribeiro-Palau, F. Lafont, J. Brun-Picard, D. Kazazis, A. Michon, F. Cheynis, O. Couturaud, C. Consejo, B. Jouault, W. Poirier and F. Schopfer, Nature Nanotechnology **10**, 965-971 (2015).
[14] C.-C. Kalmbach, J. Schurr, F. J. Ahlers, A. Müller, S. Novikov, N. Lebedeva and A. Satrapinski, Applied Physics Letters **105**, 073511 (2014).
[15] E. McCann, K. Kechedzhi, V. I. Fal'ko, H. Suzuura, T. Ando, and B. L. Altshuler, Physical Review Letters **97**, 146805 (2006).
[16] S. V. Morozov, K. S. Novoselov, M. I. Katsnelson, F. Schedin, L. A. Ponomarenko, D. Jiang, and A. K. Geim, Physical Review Letters **97**, 016801 (2006).
[17] X. Wu, X. Li, Z. Song, C. Berger, and W. A. de Heer, Physical Review Letters **98**, 136801 (2007).
[18] A. M. R. Baker, J. A. Alexander-Webber, T. Altebaeumer, T. J. B. M. Janssen, A. Tzalenchuk, S. Lara-Avila, S. Kubatkin, R. Yakimova, C.-T. Lin, L.-J. Li, and R. J. Nicholas, Physical Review B **86**, 235441 (2012).
[19] S. Feng, P. A. Lee, and A. D. Stone, Phys. Rev. Lett. **56**, 1960 (1986).
[20] P. A. Lee, A. D. Stone, and H. Fukuyama, Phys. Rev. B **35**, 1039 (1987).
[21] T. J. Thornton, M. Pepper, H. Ahmed. G. J. Davies, and D. Andrews, Phys. Rev. B **36**, 4514 (1987).
[22] J. R. Gao, J. Caro, A. H. Verbruggen, S. Radelaar, and J. Middelhoek, Phys. Rev. B **40**, 11676 (1989).
[23] F. Hohls, U. Zeitler, and R. J. Haug, Ann. Phys. **8**, Spec. Issue, 97-100 (1999).
[24] V. Skákalová, A. B. Kaiser, J. S. Yoo, D. Obergfell, and S. Roth, Phys. Rev. B **80**, 153404 (2009).
[25] G. Bohra, R. Somphonsane, N. Aoki, Y. Ochiai, D. K. Ferry, and J. P. Bird, Appl. Phys. Lett. 101, 093110 (2012).
[26] D. E. Beutler, T. L. Meisenheimer, and N. Giordano, Phys. Rev. Lett. **58**, 1240 (1987); T. L. Meisenheimer, D. E. Beutler, and N. Giordano, Jpn. J. Appl. Phys. Suppl. **26**, 295 (1987); G. A. Garfunkel, G. B. Alers, M. B. Weissman, J. M. Mochel, and D. J. VanHarlingen, Phys. Rev. Lett. **60**, 2773 (1988); T. L. Meisenheimer and N. Giordano, Phys. Rev. B 39, 9929 (1989).
[27] P. McConville, and N. O. Birge, Phys. Rev. B **47**, 16667 (1993).
[28] T. Ludwig, Ya. M. Blanter, and A. D. Mirlin, Phys. Rev. B **70**, 235315 (2004).
[29] C. Barone, F. Romeo, A. Galdi, P. Orgiani, L. Maritato, A. Guarino, A. Nigro, and S. Pagano, Phys. Rev. B **87**, 245113 (2013).
[30] C. Barone, A. Guarino, A. Nigro, A. Romano, and S. Pagano, Phys. Rev. B **80**, 224405 (2009).
[31] M. Kruskopf, K. Pierz, S. Wundrack, R. Stosch, T. Dziomba, C.-C. Kalmbach, A. Müller, J. Baringhaus, C. Tegenkamp, F. J Ahlers and H. W Schumacher, J. Phys.: Condens. Matter **27**, 185303 (2015).
[32] T. Yager, A. Lartsev, S. Mahashabde, S. Charpentier, D. Davidovikj, A. Danilov, R. Yakimova, V. Panchal, O. Kazakova, A. Tzalenchuk, S. Lara-Avila, and S. Kubatkin, Nano Lett. **13**, 4217 (2013).
[33] L. Wang, I. Meric, P. Y. Huang, Q. Gao, H. Tran, T. Taniguchi, K. Watanabe, L. M. Campos, D. A. Muller, J. Guo, P. Kim, J. Hone, K. L. Shepard, and C. R. Dean, Science **342**, 614 (2013).
[34] S. Lara-Avila, K. Moth-Poulsen, R. Yakimova, T. Bjørnholm, V. Fal'ko, A. Tzalenchuk, and S. Kubatkin, Adv. Mater. **23**, 878 (2011).
[35] J. Schurr, H. Moser, K. Pierz, G. Ramm, and B. P. Kibble, IEEE Trans. Instrum. Meas. **60**, 2280 (2011).
[36] J. Schurr, F. Ahlers, and L. Callegaro, IEEE Trans. Instrum. Meas. **62**, 1574-1579 (2013).
[37] A. A. Kaverzin, A. S. Mayorov, A. Shytov, and D. W. Horsell, Phys. Rev. B **85**, 075435 (2012).
[38] K. I. Bolotin, K. J. Sikes J. Hone, H. L. Stormer, and P. Kim, Phys. Rev. Lett. **101**, 096802 (2008).

39 N. O. Birge, B. Golding, and W. H. Haemmerle, Phys. Rev. Lett. 62, 195 (1989).
40 A. M. R. Baker, J. A. Alexander-Webber, T. Altebaeumer, S. D. McMullan, T. J. B. M. Janssen, A. Tzalenchuk, S. Lara-Avila, S. Kubatkin, R. Yakimova, C.-T. Lin, L.-J. Li, and R. J. Nicholas, Physical Review B 87, 045414 (2013).
41 S. S. Kubakaddi, Phys. Rev. B 79, 075417 (2009).
42 S. Lara-Avila, A. Tzalenchuk, S. Kubatkin, R. Yakimova, T. J. B. M. Janssen, K. Cedergren, T. Bergsten, and V. Fal'ko, Physical Review Letters 107, 166602 (2011).
43 J. Jobst, D. Waldmann, I. V. Gornyi, A. D. Mirlin, and H. B. Weber, Physical Review Letters 108, 106601 (2012).
44 T. Stauber, N. M. R. Peres, and F. Guinea, Phys. Rev. B 76, 205423 (2007); N. M. R. Peres, J. M. B. Lopes dos Santos, and T. Stauber, *ibid.* 76, 073412 (2007).
45 E. H. Hwang and S. Das Sarma, Phys. Rev. B 77, 115449 (2008).
46 J. Y. Wang, R. Q. Zhao, M. M. Yang, Z. F. Liu, and Z. R. Liu, J. Chem.Phys. 138, 084701 (2013).
47 Z. Li, J. Wang, and Z. Liu, The Journal of Chemical Physics 141, 144107 (2014).
48 A. D. Stone, Phys. Rev. B 39, 10736 (1989).
49 S. Xiong and A. Douglas Stone, Physical Review Letters 68, 3757 (1992).
50 P. A. Lee, A. Douglas Stone, and H. Fukuyama, Physical Review B 35, 1039 (1987).
51 B. L. Altshuler and B. I. Shklovskii, Zh. Eksp. Teor. Fiz. 91, 220 (1986) [Sov. Phys. JETP 64, 127 (1985)].
52 T. Machida, S. Ishizuka, S. Komiyama, K. Muraki, and Y. Hirayama, Physical Review B 63, 045318 (2001).
53 D. B. Chklovskii and P. A. Lee, Phys. Rev. B 48, 18060 (1993).
54 D. S. Lee, V. Skákalová, R. T. Weitz, K. von Klitzing, and J. H. Smet, Physical Review Letters 109, 056602 (2012).
55 J. A. Simmons, S. W. Hwang, D. C. Tsui, H. P. Wei, L. W. Engel, and M. Shayegan, Physical Review B 44, 12933 (1991).
56 J. Velasco, Jr., G. Liu, L. Jing, P. Kratz, H. Zhang, W. Bao, M. Bockrath, and C. N. Lau, Physical Review B 81, 121407(R) (2010).
57 S. Branchaud, A. Kam, P. Zawadzki, F. M. Peeters, and A. S. Sachrajda, Physical Review B 81, 121406(R) (2010).
58 S. Ilani, J. Martin, E. Teitelbaum, J. H. Smet, D. Mahalu, V. Umansky & A. Yacoby, Nature 427, 328-332 (2004).
59 J. Martin, N. Akerman, G. Ulbricht, T. Lohmann, K. von Klitzing, J. H. Smet & A. Yacoby, Nature Physics 5, 669-674 (2009).
60 S. Jung, G. M. Rutter, N. N. Klimov, D. B. Newell, I. Calizo, A. R. Hight-Walker, N. B. Zhitenev, and J. A. Stroscio, Nature Physics 7, 245-251 (2011).
61 P. C. Main, A. K. Geim, H. A. Carmona, C. V. Brown, and T. J. Foster, Physical Review B 50, 4450 (1994).
62 S. Kopylov, A. Tzalenchuk, S. Kubatkin, and V. I. Fal'ko, Appl. Phys. Lett. 97, 112109 (2010).
63 T. J. B. M. Janssen, A. Tzalenchuk, R. Yakimova, S. Kubatkin, S. Lara-Avila, S. Kopylov, and V. I. Fal'ko, Phys. Rev. B 83, 233402 (2011).
64 F. Hohls, U. Zeitler, and R. J. Haug, Physical Review B 66, 073304 (2002).
65 S. L. Rumyantsev, D. Coquillat, R. Ribeiro, M. Goiran, W. Knap, M. S. Shur, A. A. Balandin, and M. E. Levinshtein, Applied Physics Letters 103, 173114 (2013).
66 D. C. Glattli, Eur. Phys. J. Special Topics 172, 163-179 (2009).
67 B. W. Ricketts, J. Phys. D: Appl. Phys. 18, 885-892 (1985).
68 J. Schurr, H. Moser, K. Pierz, G. Ramm, and B. P. Kibble, IEEE Trans. Instrum. Meas. 60, 2280-2285 (2011).
69 K. Chida, T. Arakawa, S. Matsuo, Y. Nishihara, T. Tanaka, D. Chiba, T. Ono, T. Hata, K. Kobayashi, and T. Machida, Physical Review B 87, 155313 (2013).
70 D. G. Polyakov and B. I. Shklovskii, Physical Review Letters 70, 3796 (1993).
71 D. G. Polyakov and B. I. Shklovskii, Physical Review B 48, 11167 (1993).
72 F. Hohls, U. Zeitler, and R. J. Haug, Physical Review Letters 88, 036802 (2002).
73 J. A. Alexander-Webber, J. Huang, D. K. Maude, T. J. B. M. Janssen, A. Tzalenchuk, V. Antonov, T. Yager, S. Lara-Avila, S. Kubatkin, R. Yakimova, and R. J. Nicholas, Scientific Reports 6, 30296 (2016).
74 M. Furlan, Physical Review B 57, 14818 (1998).